\documentclass[prl,reprint,showpacs,amsmath,amssymb,superscriptaddress,longbibliography]{revtex4-1}
\usepackage{graphicx,bm,color,mathptmx,dcolumn}
\usepackage[colorlinks=true,linkcolor=blue,citecolor=blue,urlcolor =blue]{hyperref}

\newcommand{\Tr}{\mathop{\mathrm{Tr}} \nolimits}
\newcommand{\op}[1]{#1}

\begin{document}

\title{Evading Vacuum Noise: Wigner Projections or Husimi Samples?}

\author{C.~R.~M\"{u}ller}
\affiliation{Max-Planck-Institut f\"ur die Physik des Lichts,
  G\"{u}nther-Scharowsky-Stra{\ss}e 1, Bau 24, 91058 Erlangen, Germany}
\affiliation{ Institut f\"ur Optik, Information und Photonik,
 Universit\"{a}t Erlangen-N\"{u}rnberg,  Staudtstra{\ss}e 7/B2,
 91058 Erlangen, Germany}

\author{C.~Peuntinger}
\affiliation{Max-Planck-Institut f\"ur die Physik des Lichts,
  G\"{u}nther-Scharowsky-Stra{\ss}e 1, Bau 24, 91058 Erlangen, Germany}
\affiliation{ Institut f\"ur Optik, Information und Photonik,
 Universit\"{a}t Erlangen-N\"{u}rnberg,  Staudtstra{\ss}e 7/B2,
 91058 Erlangen, Germany}
\affiliation{Department of Physics, University of Otago, 730
Cumberland Street, Dunedin, New Zealand}

\author{T.~Dirmeier}
\affiliation{Max-Planck-Institut f\"ur die Physik des Lichts,
  G\"{u}nther-Scharowsky-Stra{\ss}e 1, Bau 24, 91058 Erlangen, Germany}
\affiliation{ Institut f\"ur Optik, Information und Photonik,
 Universit\"{a}t Erlangen-N\"{u}rnberg,  Staudtstra{\ss}e 7/B2,
 91058 Erlangen, Germany}

\author{I.~Khan}
\affiliation{Max-Planck-Institut f\"ur die Physik des Lichts,
  G\"{u}nther-Scharowsky-Stra{\ss}e 1, Bau 24, 91058 Erlangen, Germany}
\affiliation{ Institut f\"ur Optik, Information und Photonik,
 Universit\"{a}t Erlangen-N\"{u}rnberg,  Staudtstra{\ss}e 7/B2,
 91058 Erlangen, Germany}

\author{U.~Vogl}
\affiliation{Max-Planck-Institut f\"ur die Physik des Lichts,
  G\"{u}nther-Scharowsky-Stra{\ss}e 1, Bau 24, 91058 Erlangen, Germany}
\affiliation{ Institut f\"ur Optik, Information und Photonik,
 Universit\"{a}t Erlangen-N\"{u}rnberg,  Staudtstra{\ss}e 7/B2,
 91058 Erlangen, Germany}

\author{Ch.~Marquardt}
\affiliation{Max-Planck-Institut f\"ur die Physik des Lichts,
  G\"{u}nther-Scharowsky-Stra{\ss}e 1, Bau 24, 91058 Erlangen, Germany}
\affiliation{ Institut f\"ur Optik, Information und Photonik,
 Universit\"{a}t Erlangen-N\"{u}rnberg,  Staudtstra{\ss}e 7/B2,
 91058 Erlangen, Germany}

\author{G.~Leuchs}
\affiliation{Max-Planck-Institut f\"ur die Physik des Lichts,
  G\"{u}nther-Scharowsky-Stra{\ss}e 1, Bau 24, 91058 Erlangen, Germany}
\affiliation{ Institut f\"ur Optik, Information und Photonik,
 Universit\"{a}t Erlangen-N\"{u}rnberg,  Staudtstra{\ss}e 7/B2,
 91058 Erlangen, Germany}

\author{L.~L.~S\'{a}nchez-Soto}
\affiliation{Max-Planck-Institut f\"ur  die Physik des Lichts,
 G\"{u}nther-Scharowsky-Stra{\ss}e 1, Bau 24,  91058 Erlangen,
Germany}
\affiliation{ Institut f\"ur Optik, Information und Photonik,
 Universit\"{a}t Erlangen-N\"{u}rnberg,  Staudtstra{\ss}e 7/B2,
 91058 Erlangen, Germany}
\affiliation{Departamento de \'Optica, Facultad de F\'{\i}sica,
 Universidad Complutense, 28040 Madrid, Spain}

\author{Y.~S.~Teo}
\affiliation{Department of Optics, Palack\'{y}  University,
17. listopadu 12, 77146 Olomouc, Czech Republic}

\author{Z.~Hradil}
\affiliation{Department of Optics, Palack\'{y}  University,
17. listopadu 12, 77146 Olomouc, Czech Republic}

\author{J.~\v{R}eh\'{a}\v{c}ek}
\affiliation{Department of Optics, Palack\'{y}  University,
17. listopadu 12, 77146 Olomouc, Czech Republic}

\begin{abstract}
  The accuracy in determining the quantum state of a system depends on
  the type of measurement performed. Homodyne and heterodyne detection
  are the two main schemes in continuous-variable quantum
  information. The former leads to a direct reconstruction of the
  Wigner function of the state, whereas the latter samples its Husimi
  $Q$~function. We experimentally demonstrate that heterodyne
  detection outperforms homodyne detection for almost all Gaussian
  states, the details of which depend on the squeezing strength and
  thermal noise.
\end{abstract}

\pacs{03.65.Ta, 03.67.Hk, 42.50.Dv, 42.50.Lc}

\maketitle

\emph{Introduction.---}
Quantum information has achieved remarkable progress in the past few
years and promises even more far-reaching advances in the near future.
Pioneering proposals, such as quantum
cryptography~\cite{Bennett:1984sv,Ekert:1991qb} and
teleportation~\cite{Bennett:1993pb}, just to list the most popular,
have been demonstrated in numerous experiments. Furthermore, some of
them are already in commercial operation~\cite{idquant}.

The key concepts in the field were initially developed mainly for
discrete variables, more specifically for qubits. The
continuous-variable (CV) approach offers many practical advantages
though~\cite{Braunstein:2005aa,Ferraro:2005ns,
  CV2007:aa,Andersen:2010ng,Adesso:2014pm}. Here, information is
encoded in continuous degrees of freedom; for example, the quadratures
of a field mode. Interestingly enough, in this CV setting most
protocols can be simply implemented with linear optical components.

Gaussian states constitute a primary tool for CV quantum information
processing~\cite{Weedbrook:2012ag}. They are versatile resources
particularly easy to prepare and control. In addition, they are
completely characterized by a finite number of parameters (the
covariance matrix of the canonical mode operators), despite their
infinite-dimensional support. We shall restrict our attention to this
set of states.

To capitalize on CV resources, a very efficient
detection is paramount. In optical CV implementations, there are two
well established schemes. The first one is homodyne
detection~\cite{Yuen:1983ba,Abbas:1983ak, Schumaker:1984qm}, which
performs a projective measurement of a rotated field quadrature. This
is precisely the marginal distribution of the Wigner function of the
state~\cite{Vogel:1989zr}, that can thus be efficiently
reconstructed~\cite{Lvovsky:2009ys}. This method has been shown to
achieve the ultimate resolution predicted by the Fisher
information~\cite{Hradil:1996ta}, so it comes as no surprise that 
it is widely considered to be optimal in the CV community.

The other technique, heterodyne detection~\cite{Javan:1962aa,
  Read:1965aa,Carleton:1968aa,Gerhardt:1972aa,Yuen:1980ys,
  Shapiro:1984aa,Shapiro:1985aa,Walker:1986qn,Collett:1987aa,
  Lai:1989ah,DAriano:1997aa}, realizes the approximate measurement of
two complementary orthogonal quadratures. This corresponds to a direct
sampling of the Husimi~$Q$ function~\cite{Stenholm:1992ps}. The price
to be paid for a simultaneous measurement of noncommuting observables
is the presence of additional vacuum noise, as was first pointed out
by Arthurs and Kelly~\cite{Arthurs:1965al}.

In theory, both the Wigner and Husimi functions are equivalent
representations of the state. However, when reconstructed from
experimental data, only finite data resources are available. Technical
details aside, this fundamental limitation imposes restrictions on the
accuracy of the reconstructions.

In this Letter, we make an unbiased assessment of the accuracy of both
schemes in realistic scenarios. We shall experimentally corroborate
that for almost all Gaussian states, except for states close to the
vacuum, heterodyne detection outperforms homodyne detection. 
We emphasize that this is more than an academic curiosity, for the 
Wigner function and the Husimi function are a general concept in many fields
of physics and our results are of practical relevance in protocols 
such as CV quantum key distribution~\cite{Lorenz:2004aa, Lance:2005aa,Scarani:2009cq}.


\emph{Characterizing covariance matrices.---}
In homodyne detection, one measures the intensities at the outputs of
a beam splitter that coherently merges the signal mode and a local
oscillator. In this way, data points $x_{\theta}$ are sampled from the
marginal distribution of the Wigner function projected along a
specific field quadrature at the phase-space angle $\theta$. As we are
dealing with Gaussian states, they are fully characterized by the
covariance matrix ${G}_{\textsc{w}}$ of the Wigner function. We recall
that for a vector $\mathbf{Y} = (Y_{1}\, \ldots \,Y_{n})^{\textsc{t}}$
of random variables $Y_j$ ($\textsc{t}$ stands for the transpose), the
elements of the covariance matrix are
$G_{ij} = \mathrm{Cov} (Y_{i}, Y_{j}) = \langle ( Y_{i} - \langle
Y_{i}\rangle ) ( Y_{j} - \langle Y_{j}\rangle ) \rangle$.
In what follows, we shall set $\langle \mathbf{Y} \rangle = 0$, since the independent estimation of a trivial phase-space translation does not change the qualitative trade-off between the two measurement schemes. 


Heterodyne detection originally referred to the beating of a signal
with a slightly detuned local oscillator. Nowadays, it also comprises
two simultaneous homodyne measurements in orthogonal quadratures with
a local oscillator of the same frequency: the signal is split by a
symmetric beam splitter prior to being projected onto the quadratures
$\op{X}$ and $\op{P}$.  In both cases, data gathered for the same
phase-space angle $\theta$ are sampled directly from the Husimi
$Q$~function, which is the conditional distribution for the pair of
complementary quadratures.  As the commutation relations are
preserved, the split signals are convoluted with the vacuum
noise that can be visualized as entering the unused port of the beam
splitter.  The covariance matrix for the $Q$~function is then
${G}_\textsc{q} = {G}_\textsc{w} +   \openone$ (we normalize
the variance of the vacuum noise to unity).

Real detectors possess efficiencies $\eta<1$, so that the inferred 
covariance matrices for the two schemes are
respectively~\cite{Rehacek:2015qp} 
\begin{equation}
  {G}_\textsc{hom} = {G}_\textsc{w}+  
  \tfrac{1-\eta}{\eta} \openone \,,
\qquad
  {G}_\textsc{het} = {G}_\textsc{hom} + 
 \tfrac{1}{\eta} \openone \,.
  \label{eq:ConnectionGhomGhet}
\end{equation}

One may confidently make sense of these results from the
concepts of marginal and conditional probability distributions. Data
sampled according to these two distributions lead to rather
different uncertainties. Along the phase-space direction of the unit
vector $\mathbf{n} = (\cos \theta\,\,\sin \theta)^\textsc{t}$, the
marginal variance $\sigma^2_\theta$ (i.e., homodyne) and
conditional variance $\Sigma^2_\theta$ (i.e., heterodyne) 
are~\cite{Cox:2006dv}
\begin{equation}
  \sigma^2_\theta = 
     \mathbf{n}^\textsc{t}  \, {G}_\textsc{hom} \, \mathbf{n}\,,
  \qquad
  \Sigma^2_\theta = 
 ( \mathbf{n}^\textsc{t} \,
  {G}_\textsc{het}^{-1} \,  \mathbf{n} )^{-1}\, .
\end{equation}
If the extra noise term $\tfrac{1}{\eta} \openone$ in
${G}_\textsc{het}$ were absent, then it is easy to show that
$\sigma^2_\theta \geq \Sigma^2_\theta$ for any Gaussian state.   
The equality holds for rotationally symmetric states, as is the
case for the vacuum itself. 

\begin{figure}
  \includegraphics[width=\columnwidth]{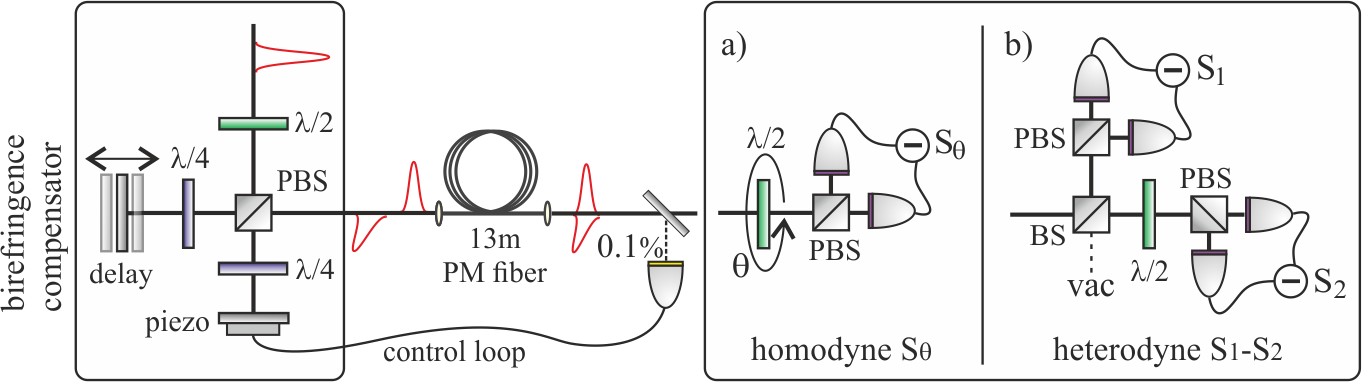}
  \caption{Sketch of the experimental setup.  Pulses from a shot-noise
    limited laser centered at 1560\,nm are distributed equally on
    the principal axes of a 13\,m long polarization maintaining fiber.
    The pulses are individually squeezed due to the Kerr
    nonlinearity. A birefringent compensator controls the temporal
    overlap of the emerging pulses and locks them to a relative phase
    difference of $\pi/2$, hence forming a $S_3$ polarized state.  The
    a)~homodyne and b)~heterodyne schemes are emulated by Stokes
    measurements.}
  \label{fig:setup}
\end{figure}

However, the additional noise $\tfrac{1}{\eta} \openone$ introduces 
further complexity. For the pertinent example of the vacuum
(${G}_\textsc{w} = \openone$), we have now $\sigma^2_\theta = 
\tfrac{1}{\eta} \leq \tfrac{2}{\eta} = \Sigma^2_\theta$, so that the 
uncertainty for data acquired from the marginal
distribution (Wigner) is less than those from the conditional
distribution (Husimi). Nonetheless, as we shall soon see, for
sufficiently bright thermal states, the additional noise in the
heterodyne detection becomes negligible, such that, above a threshold
thermal photon number, the detrimental impact of the beam splitting
noise is overcompensated by the advantage of obtaining two sample
points per signal state, ultimately rendering heterodyne detection the
superior strategy.

In the same vein, squeezing improves the tomographic performance of
heterodyne data over homodyne data, thereby surmounting the intrinsic
Arthurs-Kelly measurement uncertainties.

\emph{Experimental setup.---}
With a centered and appropriately oriented coordinate system, the
covariance matrix is completely determined by two parameters.  A
convenient representation in terms of the ellipticity $\lambda$ and
phase-insensitive noise $\mu$ yields
\begin{equation}
  {G}_\textsc{w} =  \mu
  \left (
    \begin{array}{cc}
      1/\lambda & 0 \\
      0 & \lambda
    \end{array}
  \right ) \, .
  \label{eq:CovParametrized}
\end{equation}
To check the predictions in Ref.~\cite{Rehacek:2015qp}, we prepared
states with different $\lambda$ and $\mu$ parameters employing a
fiber-based polarization squeezing setup~\cite{Heersink:2005ul}
sketched in Fig.~\ref{fig:setup}.  A shot-noise-limited laser (ORIGAMI
from Onefive GmbH) emitting 220~fs pulses at a repetition rate of
80~MHz and centered at 1560~nm is coupled equally onto the principal
axes of a 13~m-long polarization-maintaining fiber.
Quadrature-squeezed states are simultaneously and independently
generated in both polarization modes by the Kerr nonlinearity.  The
strong birefringence of the fiber causes a delay between the emerging
quadrature squeezed pulses, which is precompensated with a
Michelson-like interferometer placed before the fiber.  A weak tap-off
measurement ($\approx 0.1\%$) at the fiber output is used in a control
loop to lock the relative phase between the exiting pulses to $\pi/2$,
so that the light is circularly polarized.

\begin{figure}
  \includegraphics[width=.95\columnwidth]{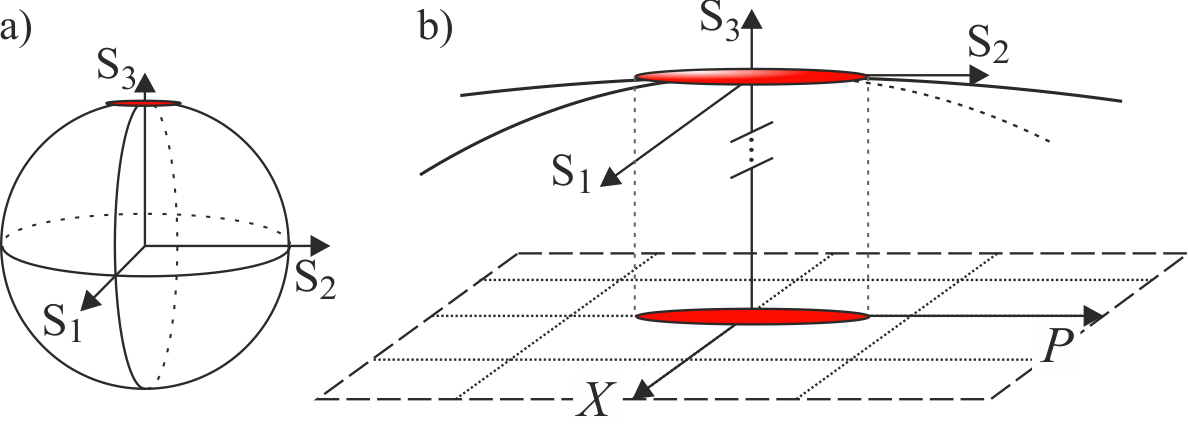}
  \caption{a) Illustration of the squeezed $S_3$-polarized state on
    the Poincar\'e sphere.  b) Magnification of the polarization state at the
    pole of the Poincar\'e sphere.  For bright states, the
    Poincar\'e sphere has a large radius such that the curvature is
    locally negligible and the projection in the $S_1$-$S_2$ dark
    plane is equivalent to a rescaled canonical $\op{X}$-$\op{P}$
    quadrature phase space.}
  \label{fig:Poincare_to_PhaseSpace}
\end{figure}

The quantum polarization of light is conveniently described by the
Stokes parameters $\op{\mathbf{S}} = (
\op{S}_{1}\,\,\, \op{S}_{2} \,\,\,\op{S}_{3})^\textsc{t}$.
For bright circularly polarized light, as generated in our setup, we
have $\langle \op{S}_{3} \rangle = \alpha^{2} \gg 1$, while
$\langle \op{S}_{1, 2} \rangle = 0$.  More generally,
$\langle \op{S}_{\theta} \rangle =0$ for the rotated version
$\op{S }_{\theta} = \op{S}_{1}\cos \theta \, + \op{S}_{2}\sin
\theta$. This defines the ``dark plane''---the plane of zero 
mean intensity. The fluctuations of these operators can then be
approximated as~\cite{Heersink:2005ul}
\begin{equation}
  \delta \op{S}_{\theta} =  \alpha ( \delta \op{X}_{\theta}  +
  \delta \op{X}_{\theta +  \pi/2} ) \, ,
  \label{eq_polsq_darkmode}
\end{equation}
where $\op{X}_{\theta}$ are the rotated dark-plane quadratures of the
bright field~\cite{Heersink:2005ul}. Such measurements are then
identical to balanced homodyne detection: the classical excitation is
a local oscillator for the orthogonally polarized dark mode, as
sketched in Fig.~\ref{fig:Poincare_to_PhaseSpace}. This is a unique
feature of polarization measurements and has been used in many
experiments~\cite{Grangier:1987fk,Josse:2004ys,
  Marquardt:2007bh,Muller:2012ys,Peuntinger:2014aa}.

Consequently, homodyne tomography is performed by sampling the
marginal distributions of the dark plane observables
$\op{X}_{\theta}$ at 100 equally separated angles
$\theta \in  [ 0, 2\pi )$. This is done by rotating a
half-wave plate positioned after the fiber output with a stepper
motor.  The heterodyne measurement is realized as simultaneous
measurements of conjugate dark plane observables $\op{X}_{\theta}$ and
$\op{X}_{\theta +\pi/2}$ (which reduce to $X$ and $P$ for $\theta = 0$).

The two detectors (InGaAs PIN photodiodes, with 98\% quantum
efficiency at DC) are balanced and provide a sub-shot noise resolution
in the frequency range between 5~MHz and 30~MHz. Each detector has two
separate outputs: DC, providing the average values of the
photocurrents, and AC, providing the photocurrents amplified in the
radio-frequency (RF) spectral range.  The RF currents of the
photodetectors are mixed with an electronic local oscillator at 12
MHz, low-pass filtered (BLP 1.9 with a --3~dB cutoff at 2.3~MHz),
amplified (FEMTO DHPVA-100), and digitized by an analog/digital exit
converter (Gage CompuScope~1610) at 10~Megasamples per second with a
16-bit resolution and 5~times oversampling.

\emph{Data Analysis and Results.---}
We experimentally prepared a vacuum state and two squeezed states with
different degrees of squeezing.  We control the squeezing strength as
well as the purity of the state by varying the pump power.  For the
strongly squeezed state, the pulse power at the fiber output was about
55~pJ (i.e., $4.33 \times10^8$ photons/pulse) and a total squeezing
strength of 2.51~dB is observed.  In the orthogonal quadrature, the
antisqueezing noise is strongly enhanced due to guided acoustic wave
Brillouin scattering~\cite{Shelby:1985ly,Shelby:1986ve,Elser:2006qf}
resulting in strongly elliptical states.  We observed a noise level of
18.78~dB above the shot noise level in the antisqueezing direction.
For the weakly squeezed state the pulse power was about 40~pJ (i.e.,
$3.15 \times 10^8$ photons/pulse) yielding a squeezing strength of
1.49~dB and a noise level of 14.62~dB in the antisqueezing direction.
The measurements are performed with different detectors and in a
temporally consecutive order.  The fluctuations of the observed
variances from their average value are within the 1\% margin.

To analyze the performance for noisy symmetric states, we
intentionally eliminate the phase information from the squeezed-state
data to emulate thermal states. The homodyne data are
\textit{thermalized} by randomly shuffling the data between the 100
measured phase bins, hence tracing out any phase information in the
data set.  For the heterodyne data, each pair of \textit{thermalized}
quadrature projections $(x, p)_{ \mathrm{therm}}$ is derived from the
measured quadrature projections $(x, p)_{\mathrm{sqz}}$ of the
squeezed states by a random rotation.

\begin{figure}[t]
  \includegraphics[width=1\columnwidth]{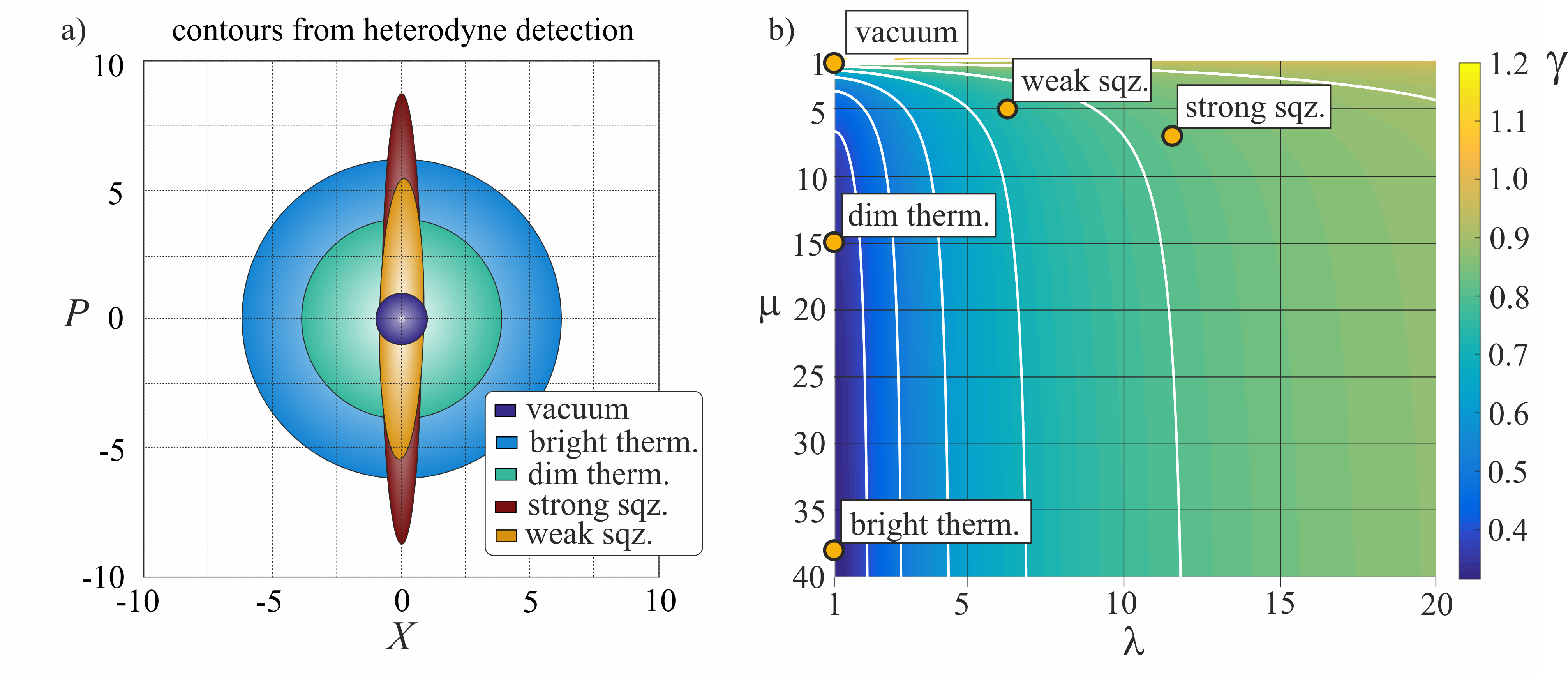}
  \caption{a)~Contours of the reconstructed covariance matrices 
    (shown here for homodyne detection). 
    b)~Color-coded contour plot of the $\gamma$ parameter in terms
    of the ellipticity $\lambda$ and thermal noise $\mu$.  The
    experimentally tested states are indicated by white disks.}
  \label{fig:Ellipses_Map}
\end{figure}

In units of coherent shot noise, the thermal states derived from the
strongly and weakly squeezed state have a (phase-invariant) quadrature
variance of 38.37 and 15.01, respectively. The variance of a thermal
state is directly proportional to the thermal mean photon number,
which yields an average thermal photon number of
$\langle N_{\mathrm{th}} \rangle= 19.2$ and
$\langle N_{\mathrm{th}} \rangle= 7.5$, respectively. The Gaussian
phase space contours of the reconstructed states are shown in
Fig.~\ref{fig:Ellipses_Map}a).

To quantify the accuracy of the estimated covariance matrix we would
need to know the true states. Since this is not strictly feasible, we
use instead the average Hilbert-Schmidt distance between the
estimation comprising a restricted number of samples ${G}_\textsc{w}$
to the estimation comprising the complete set of acquired data
$\widehat{G}_\textsc{w}$ (in our case $10^6$ homodyne samples at each
of 100 different quadrature angles and $10^8$ pairs of heterodyne
samples); that is, $ \mathbb{D}_{\mathrm{HS}} = \Tr \{ ( {G}_\textsc{w} -
\widehat{G}_\textsc{w})^2\}$. The ratio $\gamma$ between the 
Hilbert-Schmidt distances yields the relative quality of the 
reconstructions; \emph{viz} $\gamma = 
\mathbb{D}_{\mathrm{HS}}^{\scriptsize{\textsc{(het)}}}/
\mathbb{D}_{\mathrm{HS}}^{\scriptsize{\textsc{(hom)}}}$.
This ratio is essentially that of the Cram{\'e}r-Rao bounds of the two
schemes for such a large sample if $\widehat{G}_\textsc{w}$ are
maximum-likelihood (ML) estimators \cite{Rehacek:2009ys}.  Heterodyne
detection is more efficient if $\gamma<1$.  In
Fig.~\ref{fig:Ellipses_Map}b), we present a $\mu$-$\lambda$ diagram
with both the theoretical predictions and the experimental
states.

\begin{figure*}
  \includegraphics[width=2.1\columnwidth]{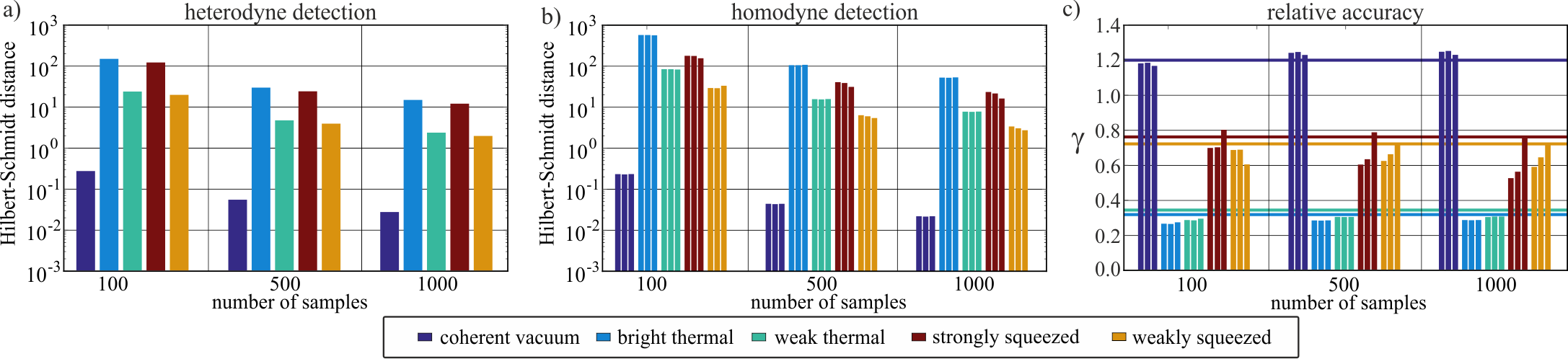}
  \caption{a) Experimental results for the accuracy of the covariance
    matrix reconstruction with heterodyne detection. The
    Hilbert-Schmidt distance is shown for three different numbers of
    sampling events.  b) Experimental results for the accuracy of the
    covariance matrix reconstruction based on homodyne detection.  c)
    Ratio of the Hilbert-Schmidt distances $\gamma$.  The solid lines
    in the background indicate the theoretical prediction. }
  \label{fig:HSD_HET_results}
\end{figure*}

\begin{table}[b]
\caption{Covariance parameters for both homodyne and heterodyne
  detection.}
\begin{ruledtabular}
\begin{tabular}{l d d  d d}
 & \multicolumn{2}{c}{heterodyne} & \multicolumn{2}{c}{homodyne}  \\
\cline{2-3} \cline{4-5}
State &  \mu  & \lambda &  \mu  & \lambda  \\
\hline
vacuum state  &  0.99 & 1.01  &  1.01 & 1.00 \\
strongly sqz. &   6.54 & 11.67 &  6.44 & 11.61 \\
weakly sqz.  &  4.62 & 6.29 &  4.46  & 6.49  \\
bright thermal &  38.36 & 1.00 &  38.40 & 1.00  \\
dim thermal  & 15.01 & 1.00 & 14.96 & 1.00
\end{tabular}
\end{ruledtabular}
\label{table:CovarianceParametersHeterodyne}
\end{table}

In the following we do not compensate for the finite efficiency of our
detectors, but  assume $\eta$ = 1 for the evaluation of the
experimental data. This is a conservative assumption, as the
superiority of heterodyne performance even increases for
$\eta<1$~\cite{Rehacek:2015qp}.

To obtain the correct $\gamma$ values, the reconstructions of
covariance matrices from the homodyne tomography data is performed via
the Gaussian Maximum Likelihood (ML) algorithm described in
Ref.~\cite{Rehacek:2009ys}.  In this algorithm, the variances of the
data collected for fixed quadrature phases are calculated and fed into
a recursive loop that optimizes the estimation for the covariance
matrix. The results are given in
Table~\ref{table:CovarianceParametersHeterodyne}.

For the heterodyne measurement, we directly calculate the sample covariance
matrix from the measured joint probability distribution, which is an ML estimator.
The data is affected by the electronic noise floor of the detectors, and by
background light which we individually record for each measurement.
We compensate for this additional noise by subtracting its covariance
matrix from the measured covariance matrix of the quantum state
${G}_\textsc{hom/het} \mapsto {G}_\textsc{hom/het} - {G}_\textsc{elec}$.
The covariance matrix of the input state is obtained then as in
Eq.~\eqref{eq:ConnectionGhomGhet}. The parameters of the
states reconstructed by heterodyne detection are also summarized in
Table~\ref{table:CovarianceParametersHeterodyne}.

To provide a fair comparison of the covariance matrix
reconstructions, the number of sampling events needs to be equal. For the
heterodyne detector one sampling event corresponds to a pair of conjugate
quadrature projections $(X, P)$, while that from the homodyne
detector is given by the single quadrature value $x_{\theta}$
projected onto the field quadrature rotated by an angle $\theta$ with
respect to an arbitrary but fixed reference frame. The reference
states, against which the accuracy of the covariance matrix estimations
are assessed, are reconstructed from extensive measurements comprising
$10^8$ samples ($100\times 10^6$ samples at different quadrature
angles for the homodyne detection). Given a fixed number of sampling events,
we still have the freedom to choose the number of quadrature
phases for the homodyne tomography. We compare the performance for 5,
10, and 15 different angles taken from the 100 angles measured in the
extensive reference measurement. To avoid a bias towards a preferred
reference frame, these angles are chosen randomly but with constant
phase difference between neighboring angles (e.g., 
$ \Delta\theta = 2\pi/5$ for the 5 angles).

The experimentally observed Hilbert-Schmidt distances for both
detection strategies as well as the corresponding $\gamma$ parameter
for various sample sizes and for different numbers of measurement angles
are shown in Fig.~\ref{fig:HSD_HET_results}. The measured $\gamma$
is in extremely good agreement with the theoretical prediction.

In agreement with theoretical predictions, the Hilbert-Schmidt
distances of both the homodyne and the heterodyne detection decrease
by about an order of magnitude when the number of sampling events is
increased by a factor of 10. As could be expected, the estimation
accuracy of the phase-covariant states (vacuum, bright/weak thermal)
is independent of the number of quadrature angles. For elliptical
states, however, the Hilbert-Schmidt distance is initially decreasing
with the number of angles, i.e. the estimation gets more accurate by
increasing the number of quadrature phases. This is particularly
pronounced for the bigger sample sizes, for which the theoretically
$\gamma$ parameter is only approached with increasing number of
quadrature phases. However, if for a given number of samples too many 
angles are measured the statistical uncertainty per quadrature is getting 
worse and the accuracy eventually decreases again. 

We also performed Gaussianity tests on the experimental data to check
for any non-Gaussian feature. For this, we sort the data into
histograms of 101 bins and use the Kolmogorov--Smirnov test and the
$\chi^2$ tests, as well as the Kullback--Leibler
divergence~\cite{Thode:2002cr}. We find that all data sets are
Gaussian within confidence levels ranging from 95 to 99\%.

\emph{Concluding remarks.---}
Apart from a small region of states close to the vacuum state,
heterodyne detection outperforms homodyne detection in terms of the
reconstruction accuracy for almost all Gaussian states.  Therefore, at
least for Gaussian states, our direct experimental confirmation of the
theory for the performances of the two detection schemes rigorously
shows the advantage of the heterodyne detection for quantum
tomography.  On this note, we believe that this result would be
particularly appealing to experimentalists, as the heterodyne
detection is conveniently realized without the need for active phase
changing elements of the local oscillator beam. Especially in CV
quantum key distribution covariance estimation is a crucial part.
Moreover, as heterodyne detection directly samples the Husimi
$Q$~function, the need for time-consuming tomographic reconstruction
is omitted. The extension of these results to non-Gaussian states is
the subject of further research.

\emph{Acknowledgments.---}
We thank Herbert Welling and Sascha Wallentowitz for discussions about
different aspects of optical heterodyne detection.  We acknowledge
financial support from the European Research Council (Advanced Grant
PACART), the Spanish MINECO (Grant FIS2015-67963-P), the Technology
Agency of the Czech Republic (Grant TE01020229), the Grant Agency of
the Czech Republic (Grant No. 15-03194S), and the IGA Project of the
Palack{\'y} University (Grant No. IGA PrF 2016-005).

\end{document}